\documentclass{article}
\usepackage{spconf,amsmath,graphicx}
\usepackage{multirow}
\usepackage{makecell}
\usepackage{booktabs}

\usepackage{amssymb}
\usepackage{amsmath}

\title{PROBING DEEP SPEAKER EMBEDDINGS FOR SPEAKER-RELATED TASKS}
\name{Zifeng Zhao$^{1,\dag}$, Ding Pan$^{1,\dag}$, Junyi Peng$^2$, Rongzhi Gu$^{3,*}$
\thanks{\dag indicates equal contribution.}
\thanks{*Corresponding author: lorrygu@tencent.com}
}
\address{
  $^1$ School of ECE, Peking University, China \\
  $^2$ Faculty of Information Technology, Brno University of Technology, Czechia \\
  $^3$ Tencent AI Lab, China
}
\begin{document}
%
\maketitle
\begin{abstract}

Deep speaker embeddings have shown promising results in speaker recognition, as well as in other \textit{speaker-related tasks}. However, some issues are still under explored, for instance, the information encoded in these representations and their influence on downstream tasks. Four deep speaker embeddings are studied in this paper, namely, d-vector, x-vector, ResNetSE-34 and ECAPA-TDNN. Inspired by human voice mechanisms, we explored possibly encoded information from perspectives of \textit{identity}, \textit{contents} and \textit{channels}; Based on this, experiments were conducted on three categories of speaker-related tasks to further explore impacts of different deep embeddings, including \textit{discriminative tasks} (speaker verification and diarization), \textit{guiding tasks} (target speaker detection and extraction) and \textit{regulating tasks} (multi-speaker text-to-speech). Results show that all deep embeddings encoded channel and content information in addition to speaker identity, but the extent could vary and their performance on speaker-related tasks can be tremendously different: ECAPA-TDNN is dominant in discriminative tasks, and d-vector leads the guiding tasks, while regulating task is less sensitive to the choice of speaker representations. These may benefit future research utilizing speaker embeddings.

\end{abstract}
\begin{keywords}
Speaker embeddings, speaker recognition, speaker-related tasks
\end{keywords}

\section{Introduction}
\label{sec:intro}

    
    \begin{figure}[ht]
      \centering
      \includegraphics[width=8cm]{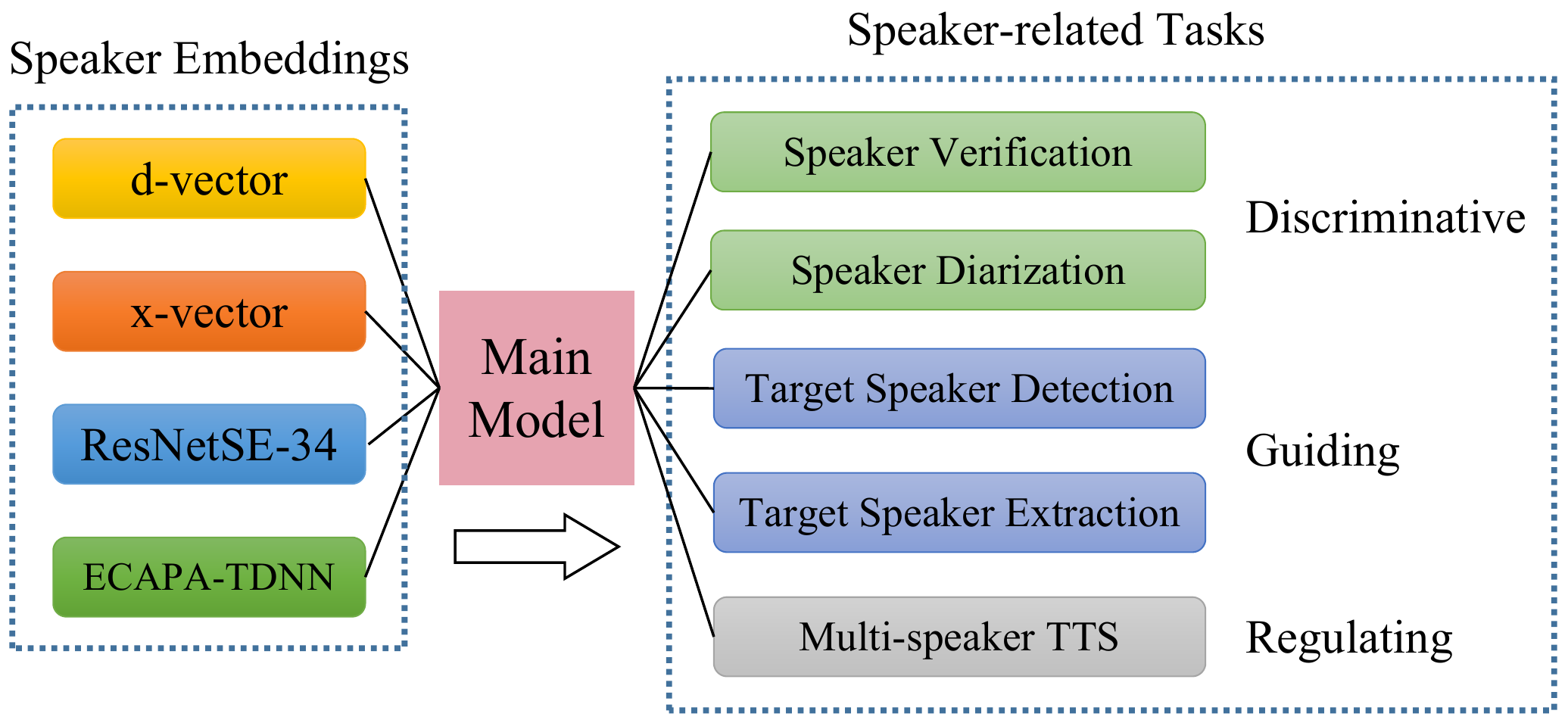}
      \caption{Speaker embeddings serve as bias or indicators for various personalized and customized speech processing tasks.}
      \label{fig:downstreams}
      \vspace{-2em}
    \end{figure}
    
Speaker recognition has made substantial progress in the past two decades. From factor analysis to deep learning methods, modern speaker recognition systems are generally built based on \textit{speaker embeddings}, for example, i-vector and x-vector. In addition to speaker recognition, speaker embeddings are also frequently used in various \textit{speaker-related tasks}, where embedding vectors are used as identity representations to encode speaker information, in order to promote personalized and customized speech processing. In multi-speaker text-to-speech, for instance, speaker embeddings help the acoustic model to generate high-quality speech with timbre similar to the enrolled speaker; And in target speaker detection, speaker embedding acts as a clue for detecting the target speaker \cite{PersonalVAD}.

Previous researches, however, show that speaker embeddings also encoded other information besides the desired speaker identity \cite{what-does}\cite{probing}\cite{empirical-analysis}. This may affect the performance of speaker recognition as well as other speaker-related tasks. In this paper, four deep speaker embeddings are studied, namely, d-vector, x-vector, ResNetSE-34 and ECAPA-TDNN, all of which are commonly used in recent researches. To probe the information they contained, we analyzed \textit{identity}, \textit{content} and \textit{channel} factors, which are tightly correlated with source excitation, vocal tract and speech transmission in human voice generation. Based on these, we further compared their impacts on downstream speaker-related tasks and the resulting performance differences. As shown in Figure \ref{fig:downstreams}, five speaker-related tasks were examined, which can be summarized into three categories: \textit{discriminative tasks}, including speaker verification (SV) and speaker diarization (SD); \textit{guiding tasks}, including target speaker detection (TSD) and target speaker extraction (TSE); and \textit{regulating task}, including multi-speaker text-to-speech (MS-TTS). Experimental results indicate that all deep embeddings encoded channel and content information besides the desired speaker identity, but the amount could vary from one to another. Their performance on downstream speaker-related tasks can also be different: ECAPA-TDNN performs best in discrminative tasks, and d-vector is superior over others in guiding tasks, while all embedding vectors achieved similar results on regulating task. We argue that these may provide a reference for future research using speaker embeddings.

\section{Deep Speaker Embeddings}
\label{sec:spkemb}

Four deep speaker embeddings are considered in this paper:

\noindent\textbf{d-vector}. This is an early speaker representation fully based on deep neural networks (DNN). The d-vector encoder is a light-weight model trained with GE2E loss \cite{GE2E}: it consists of only 3 layers of LSTM, and a fully connected layer or attentive pooling transforms the outputs to the final embedding\footnote{https://github.com/yistLin/dvector}. 

\noindent\textbf{x-vector}. With the rising of deep learning, x-vector was proposed for text-independent speaker verification \cite{xvector}. It handles short-term temporal context using a time-delay neural network (TDNN) to extract frame-level features, which are then aggregated through a temporal statistics pooling (TSP). The aggreated utterance-level features are further processed by two hidden layers to produce the final speaker embedding. Implementation by the SpeechBrain \cite{SpeechBrain} toolkits is used. 

\noindent\textbf{ResNetSE-34}. The residual network is another popular trunk architecture. The ResNet-34 was modified to adapt to spectrogram inputs and acts as a frame-level feature extractor, with attentive statistics pooling (ASP) as the aggregation module. ResNetSE-34 \cite{MetricLearning}, or r-vector, achieves outstanding performance on several benchmarks for speaker recognition. For reproducibility, we used the model from the Sunine toolkits\footnote{https://gitlab.com/csltstu/sunine}.

\noindent\textbf{ECAPA-TDNN}. ECAPA-TDNN significantly promoted the performance of speaker recognition by introducing several modifications \cite{ECAPA-TDNN} to the TDNN architecture: (1) Res2Net with squeeze-and-excitation blocks; (2) Channels- and context-attentive statistics pooling; and (3) Multi-layer feature aggregation. Model from  SpeechBrain \cite{SpeechBrain} is used.

\section{Encoded Information}
\label{sec:prob}

\begin{figure}[th]
      \centering
      \includegraphics[width=8cm]{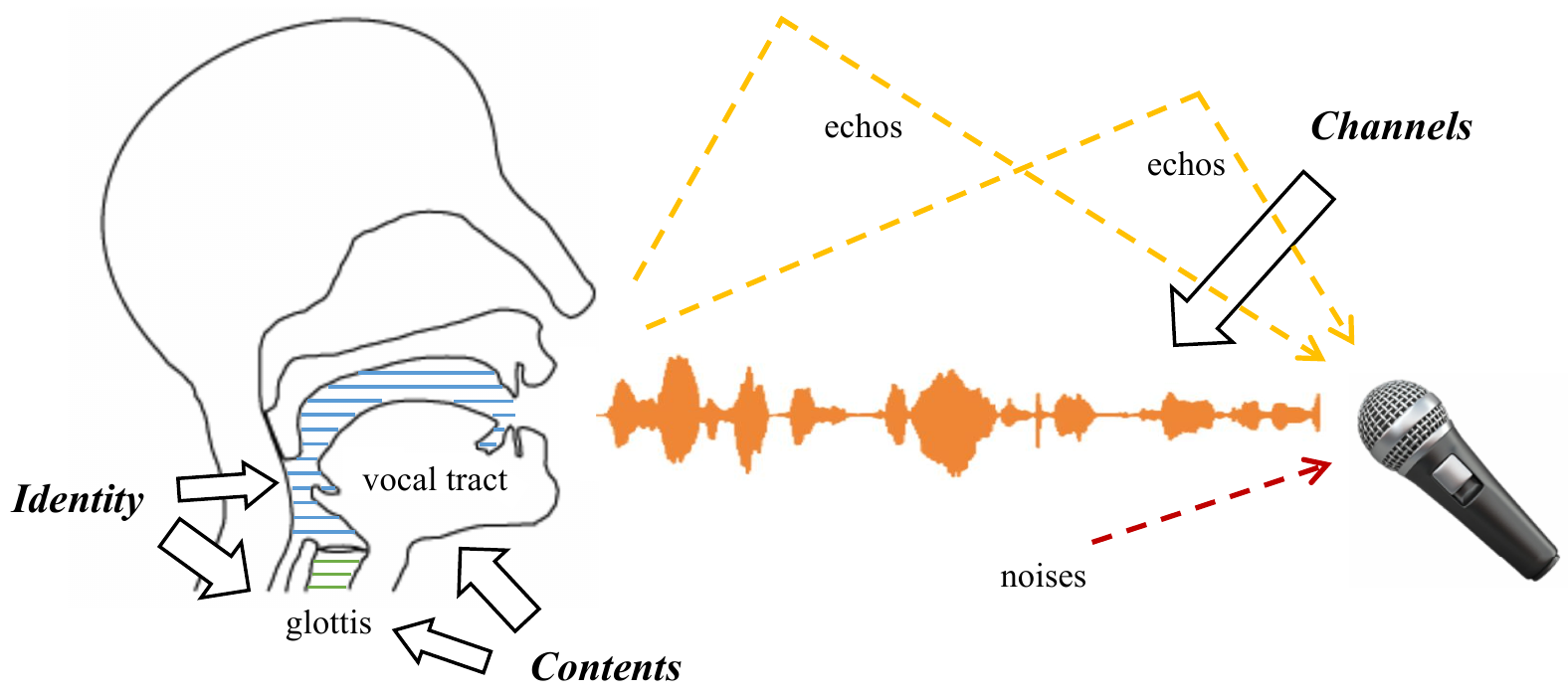}
      \caption{Speech mechanisms and conveyed information.}
      \label{fig:mechanisms}
      \vspace{-1em}
    \end{figure}

It is known that speaker embeddings contain various information besides speaker identity \cite{what-does}\cite{probing}\cite{empirical-analysis}, even though they were designed in a way to identify speakers. As depicted in Figure \ref{fig:mechanisms}, the generation of voice can be generally divided into three parts \cite{SpeechProductionTheory}: source excitation, vocal tract and speech transmission. In excitation part, airflow from the lung is controlled by the quick switching of glottis. This forms a quasi-periodic waveform, which is crucial for \textit{identifying} speakers; Periodic pulses are further modulated by the vocal tract and generate intelligible speech, which now conveys \textit{content} information as well as more detailed \textit{identity} features. Finally, the speech is transmitted in the air and affected by environment's \textit{channel} characteristics, until it is finally picked up by a device. 

Inspired by this, we summarize possibly encoded information into identity, contents and channels. Probing tasks are conducted by building simple classifiers to predict specific properties based on given speaker embeddings, where embedding vectors act as input features and few fully connected layers are used to shield useless information while retain relevant ones. Models are trained with cross-entropy loss and the classification accuracy can be considered as a proxy of how much related information was encoded in embeddings \cite{what-does}. Since identity information has been well studied in previous researches, we focus on other aspects in the following.
    
\noindent{\textbf{Channels}} Channel characteristics are determined by the acoustic environments where the speaker and microphone are located in. Relevant factors include reverberation, noises and even recording devices. For more detailed analysis, we probe channel information from two levels: scene and session.
    
\begin{figure}[th]
      \centering
      \includegraphics[width=8cm]{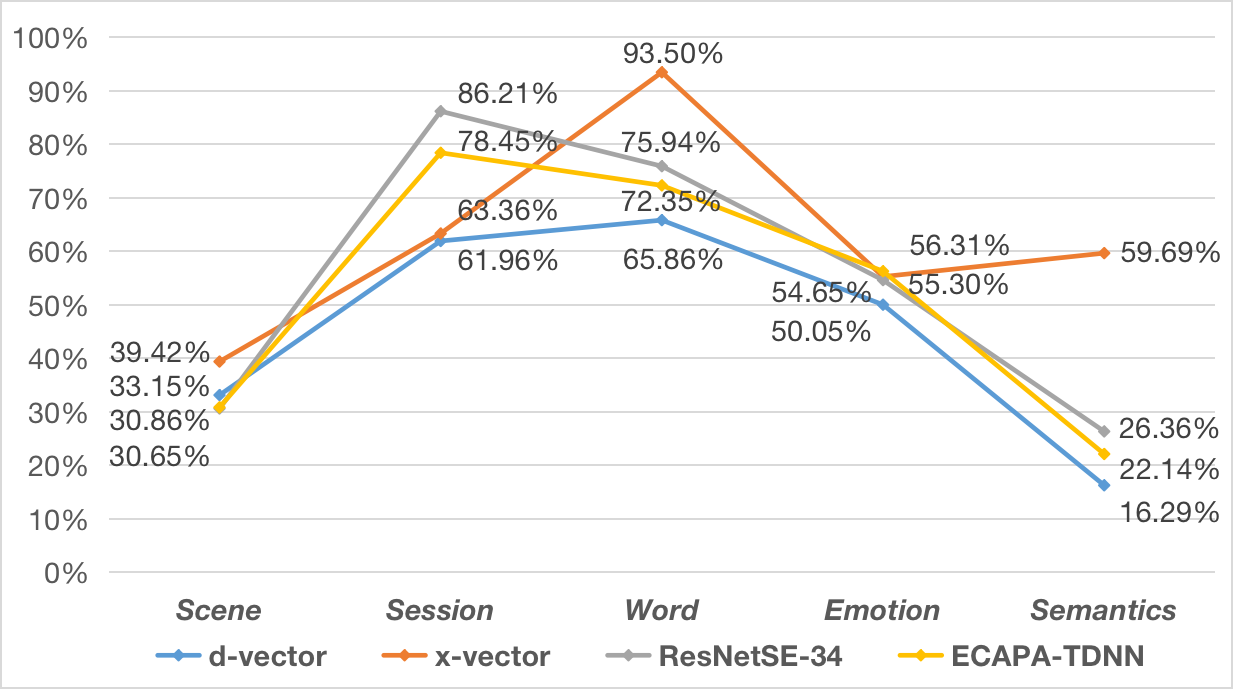}
      \caption{Classification accuracy of information probing tasks.}
      \label{fig:probing}
      \vspace{-0.5em}
    \end{figure}

Different recordings of the same \textit{scene} have similar acoustic characteristics. For instance, drama mostly happens in theatres, where reverberations are well designed while noises are strictly inhibited. Another example is vlog, which is likely to be shot by hand-held devices in the wild. We probed scene information using eleven-class scene labels provided in CN-Celeb \cite{CN-Celeb}. As shown in Figure \ref{fig:probing}, scene information is limited in all embeddings, with classification accuracy all lower than 40\%. It is safe to say that these deep embeddings contain little or limited scene-level channel information.


    
\textit{Session} is a more fine-grain channel characteristic. Even under the same scene and for same speakers, recordings may differ due to specificities in surrounding layouts, devices and acoustic environments, e.g. interviews in different rooms. We probed session information using the test set of VoxConverse \cite{VoxConverse}, which consists of 232 different session recordings.  As illustrated in Figure \ref{fig:probing}, all embeddings show high accuracy, among which ResNetSE-34 ranked top (86.21\%). Session information brings negative effects for downstream tasks: in \textit{discriminative tasks}, recordings of different speakers but with similar session characteristics may result in false accept in SV or contribute to confusion item in SD's DER metric; while for \textit{guiding tasks}, session factors can bring about a mismatch between enrollment speech and the target speech, and thus lead to \textit{target confusion problem} \cite{confusion}; And in \textit{regulating tasks}, session information hinders TTS model from precisely capturing enrolled speaker's vocal characteristics, and thus reducing speaker similarity between enrolled and the synthesized speech or even introducing artifacts.


\noindent{\textbf{Contents}} Quasi-periodic impulses from glottis are further modulated by the vocal tract and finally transmitted by the mouth, forming intelligible speech, which is a carrier of spoken language. The conveyed content information is another important aspect of speech and DNN models tend to encode it in embedding vectors. We probed three levels of speech contents, namely: word, emotion and semantics. 

\textit{Word}-level information relates to the text or lexical contents of speech signals. 20 keywords from GSCDv1 \cite{GSCD} are used for word classification to probe information. As depicted in Figure \ref{fig:probing}, all speaker embeddings show high accuracy on word classification. Most surprisingly, the performance of x-vector even reached 93.50\%, which is very close to state-of-the-art keyword spotting results. These suggest that speaker embeddings encoded large amount of word-level information, which can bring influence to downstream tasks: in \textit{discriminative tasks}, overlap on vocabulary can cause false accept, for instance, in text-independent SV; For \textit{guiding tasks}, sensitivities to words may lead to a wrong pairing between the enrollment utterance and the undesired interfering signals, especially in conversations where speakers have similar voices or the target speech is heavily corrupted.

Speech \textit{emotion} is another aspect of spoken contents. Compared with word-level contents, emotion information is mostly expressed by the prosody of speech signals. We probe emotion-level contents by a speech emotion recognition (SER) task with IEMOCAP \cite{IEMOCAP}, where expressive recordings are classified into 5 sentiments: happy, angry, sad, frustrated and neutral. Results are plotted in Figure \ref{fig:probing}. Performances range from 50.05\% to 56.31\%, with insignificant differences. The accuracies are not high, though, it should be noted that even well designed SER models hardly exceed 80\%. From this point of view, we argue that these speaker embeddings encoded a certain amount of sentimental contents. Its potential impacts on speaker-related tasks include: emotion swings between enrollment utterance and the test speech may result in a bias (e.g. a F0 shift), and hence lead to a false reject in speaker recognition; In \textit{regulating tasks}, prosody of the enrolled speech will be directly transferred to the synthesized voice via the embedding vector, for example, affecting the stress and speed; And in \textit{guiding tasks}, this is an \textit{embedding bias} as suggested in \cite{confusion}.

\textit{Semantic} information is a more abstract level of contents, focusing on the purpose and intention of the spoken contents. This is the most complex and abstract among three aforementioned content information. Probing experiments were conducted based on the Fluent \cite{Fluent} dataset, in which models were required to fill intention slots in the format of $(action,object,location)$. Those keywords may be absent in some utterances and the model needs to infer them, which makes it different from naive word recognition. All models perform flat with accuracy lower than 30\%, so it can be confidently stated that they contained very limited semantics-level information, except for x-vector (59.69\%). 

\section{Downstream Speaker-related Tasks}
\label{sec:downstream}

In previous researches, speaker embeddings are commonly used in various \textit{speaker-related tasks} to promote personalized or customized speech processing, while the impacts of different speaker encoders and the resulting performance differences are rarely considered. In this section, five speaker-related tasks are studied, so as to reveal the influence of using these speaker embeddings on different kinds of applications. Experimental results are summarized in Table \ref{tab:downstream}.

\noindent\textbf{Speaker verification (SV).} A general SV system is composed of a speaker encoder which maps the enrollment utterance and test utterance into corresponding speaker embeddings, and a back-end module which scores their similarities and judges whether they are from the same speaker. Experiments were conducted on VoxCeleb \cite{VoxCeleb1}: speaker embeddings are directly used and cosine similarity is set as the scoring back-end. ECAPA-TDNN shows superior performance with an equal error rate (EER) of 0.89\%, ResNetSE-34 ranks the second, while x-vector and d-vector performs inferior. The performance gap between  the former two and the latter two can be attributed to model capacity, and the advantage of ECAPA-TDNN over ResNetSE-34 is consistent with observations in Section 5 that ECAPA-TDNN generally encodes less unrelated information than ResNetSE-34. 

\noindent\textbf{Speaker diarization (SD).} Diarization systems are often complex, consisting of various modules including front-end processing, voice activity detection (VAD), speaker representation, clustering and so on. We mainly focus on the speaker representation part. Performance differences are compared on VoxConverse \cite{VoxConverse}. To make it simple, variable-length speech chunks after VAD are directly fed into speaker encoders without extra segmenting, followed by a spectral clustering and K-means algorithm to group them according to their embedding vectors. In Table \ref{tab:downstream}, results show similar trend with that of SV task, except for the rank between x-vector and d-vector. 

We classify SV and SD as \textit{discriminative tasks}, in which models heavily rely on the distinctiveness of embedding vectors to discriminate speakers. In both tasks, ECAPA-TDNN shows consistent superiority over others, which suggests that it is more suitable for \textit{discriminative tasks}.

\begin{table*}[t]
    \centering
    \begin{tabular}{@{}lcccccc@{}}
    \toprule
                & \multicolumn{1}{l}{SV (EER $\downarrow)$}  & \multicolumn{1}{l}{SD (DER$\downarrow)$} & \multicolumn{1}{l}{TSD (mAP$\uparrow$)} & \multicolumn{1}{l}{TSE (SI-SDRi$\uparrow$)} & \multicolumn{1}{l}{MS-TTS (MOS$\uparrow)$} & \multicolumn{1}{l}{MS-TTS (DMOS$\uparrow$)} \\ \midrule
    d-vector    & 14.75\%                      & 21.03\%                      & \textbf{0.91}                             &       \textbf{10.22}                    &          4.13                     &        4.44                        \\
    x-vector    & 3.20\%                       & 24.50\%                      & 0.77                              & 9.30                          &       3.89                           &       4.31                        \\
    ResNetSE-34 & 1.49\%                       & 18.98\%                      & 0.85                              & 7.79                          &      4.00                      &          4.56                    \\
    ECAPA-TDNN  & \textbf{0.89\%}                       & \textbf{18.37\%}                      & 0.81                              & 5.36                          &          3.81                   &       4.25                       \\ \bottomrule
    \end{tabular}
    \centering
    \caption{Comparing deep speaker embeddings on downstream speaker-related tasks}
    \label{tab:downstream}
\end{table*}

\noindent\textbf{Target speaker detection (TSD).} The goal of TSD is to detect the timestamps of target speaker's voice given his/her enrollment utterances. For this task, Embedding-VAD is proposed, which integrated aforementioned speaker embeddings with PersonalVAD \cite{PersonalVAD}: the embedding vector of enrollment speech is pre-computed, then concatenated with Fbank input and further processed by a light-weight LSTM-based model for a frame-level ternary classification. Best performance on LibriSpeech \cite{LibriSpeech} is achieved by d-vector, presumably due to that it encoded less interfering information than others. 

\noindent\textbf{Target speaker extraction (TSE).} As a variant of speech separation, TSE separates only target speaker's speech and filters out all other interferences. Generally, a TSE model is composed of a speaker encoder and a separation network. The speaker encoder encodes the enrollment speech into an embedding vector, based on which the separation network focus only on the target speaker and masks irrelevant information in the feature space. Embedding-TasNet is proposed based on Conv-TasNet \cite{Conv-TasNet}: speaker embeddings are first zero-padded to the dimension of 512, then fused with hidden representations at the end of the first dilated convolutional block. In Table \ref{tab:downstream}, there are significant performance differences when different embedding vectors are applied: d-vector ranks the top, with a SI-SDRi of over 10 dB. In contrast, ECAPA-TDNN performs very poor even though it is a strong speaker encoder.

We define TSD and TSE as \textit{guiding tasks}, since both of them use enrolled speaker embedding as a target guidance. Although inferior in speaker recognition tasks, d-vector shows significant advantages in both TSD and TSE over other embeddings, which can be attributed to less interfering information encoded, as discussed in Section 4.

\noindent\textbf{Multi-speaker text-to-speech (MS-TTS).} 
The goal of MS-TTS is to build a text-to-speech (TTS) system that can generate natural speech for a variety of speakers. A MS-TTS model normally consists of a speaker encoder and a text-to-speech network. We proposed Embedding-VITS by combining aformentioned speaker embeddings with the VITS \cite{VITS} model: pre-computed embedding vectors are directly added to the input latent variable as well as the input of stochastic duration predictor. Results show that performance difference is minor, especially the differential mean opinion score (DMOS), which measures the speaker similarity of the enrolled and generated speech. We consider MS-TTS as a \textit{regulating tasks}, in which speaker embedding encodes key speech factors to regulate the generation of acoustic features so as to synthesize speech similar to the enrolled speaker.



\noindent\textbf{Discussion.} Based on above results and analysis, we now further discuss the similarities and differences of deep speaker embeddings considered in this paper: (1) On encoded information, d-vector, ResNetSE-34 and ECAPA-TDNN show very similar trends but with different extents. X-vector performs quite abnormal in session and semantics, though, it is still consistent with others on remaining items. (2) For \textit{channel} information, all embeddings encoded limited scene-level factors, probably due to the richness of data source; While experiments on recordings classification implies that session-level factors are not negligible. (3) For \textit{content} information, the inevitable coupling of spoken contents and speaker characteristics in human speech production results in a surprisingly high sensitivity on lexical contents; \textit{Emotion} information is also entangled in these embedding vectors, while semantic information is largely restrained due to its abstractness and complexness. (4) We intuitively summarized different speaker-related tasks into three categories according to how they utilize the speaker embeddings for downstream tasks, namely, discriminative tasks, guiding tasks and regulating tasks. Results show that performance trends within the same category are similar. (5) ECAPA-TDNN is more superior on discriminative tasks thanks to less irrelevant factors encoded compared with ResNetSE-34. These two embeddings achieved outstanding performance on \textit{discriminative tasks}, though, such an advantage is not maintained on other tasks; Surprisingly, d-vector significantly outperformed others on \textit{guiding tasks}, even though it is far from state-of-the-art in speaker recognition; All performances are close on \textit{regulating tasks}, which implies that regulating task could have much weaker dependency on speaker embeddings. 

\section{Conclusion}

Four deep speaker embeddings are carefully studied in this paper, namely, d-vector, x-vector, ResNetSE-34 and ECAPA-TDNN. We first probe the information they encoded from perspectives of channels and contents, based on which we compared and analyzed their performance differences on various kinds of speaker-related tasks. Results show that ECAPA-TDNN is dominant in discriminative tasks while light-weight d-vector performs best for guiding tasks, and in regulating tasks all embeddings are close. In future work, we plan to further explore how encoded information may change when speaker encoders are jointly tuned in downstream tasks.

\newpage




\bibliographystyle{IEEEbib}
\bibliography{strings,refs}

\end{document}